\def\ket#1{|#1\rangle}

\def\Schrodinger{Schr\"{o}dinger }
\documentstyle[12pt]{article}
\begin{document}
\vskip.2in
\centerline{{\bf A  Dynamical Reduction Theory of Einstein-Podolsky-Rosen}}
\centerline{{\bf  Correlations and a Possible Origin of CP Violations.}}
\vskip.1in
\centerline{{\bf Daniel I. Fivel}}
\vskip.1in
\centerline{{\bf  Department of Physics}}
\centerline{{\bf University of Maryland, College Park, MD 20742}}
\vskip.1in
\centerline{{\bf October 12, 1996}}
\vskip.2in
\centerline{{\bf Abstract}}
\vskip.2in
We show that there is essentially only one way to construct a  stochastic
\Schrodinger equation that gives a dynamical account of the transformation of
entangled into factorized states and is consistent both with quantum mechanics and
required symmetries. The noisy, non-linear term is
 a unimodular scalar multiple of the time reversal operator that must be present whenever 
a  Hamiltonian term in the \Schrodinger equation can distinguish the factorized constituents
of an entangled state. The dynamical mechanism involved in the transformation of entangled into
 factorized states
provides an explanation for the fact that Einstein-Podolsky-Rosen correlations appear in a time
 determined by the response of the measuring device and  independent of the distance between
the particles. The dependence on the response time of the measuring device may be testable 
through a delay in observing the collapse of mesoscopic `` \Schrodinger cat" states in
ion traps. It is further shown that there are  situations  where a two-particle interaction can 
induce a non-linear term by virtue of coupling to decay modes that distinguish factorized
constituents of an entangled state. We show that this should happen
in the neutral K-meson system where the entangled $K_L$ state is pushed slightly in the
direction of a factorized constituent ($K_o$ or $\overline{K_o}$)  as a consequence of the
fact that these can be distinguished via the sign of the charged lepton in a semi-leptonic
decay mode. The result is a CP violation that is within 20\% of the experimental value.
\vskip.25in
\centerline{{\bf 1. Introduction}}
\vskip.2in
Although the most vexing conceptual problems in the foundations of quantum mechanics arise
from the manner in which the theory deals with the measurement process,
our computations are normally insensitive to these problems. We collapse
wave functions with a pencil and paper, confident that we are describing  situations
in which no future interference is possible\cite{PEARLEA}. But a small group of
physicists have always had the uneasy feeling that they were missing something interesting by sweeping the collapse
process under the rug. That group is now becoming larger as developments in
mesoscopic physics such as the study of single-atom wave functions  with trapped ions make
direct experimental probes possible.
\vskip.1in
A {\it dynamical
reduction theory} (DRT), is an extension of quantum mechanics that attempts to {\it account} for the collapse of
the wave function. DRT has a long history dating back to the earliest days of quantum mechanics. But serious conceptual
problems (such as Schr\"{o}dinger's cat) discouraged its development. Only in recent years
has it become clear that we can turn the
\Schrodinger equation into a  stochastic differential equation with a non-linear noise term that will reproduce the predictions of
quantum mechanics while collapsing wave functions rapidly in the manner that we observe.
The resulting enthusiasm is revealed by J.S. Bell's remark in 1990
 that the stochastic modification of quantum mechanics is the most 
important new idea in the foundations of quantum mechanics in his professional lifetime\cite{SHIMONY}.
\vskip.1in
Although DRT's have a non-linear term, they are not vitiated by the criticisms applicable to 
non-linear forms of the {\it non}-stochastic \Schrodinger equation, namely Peres'\cite{PERES} demonstration of
violations of the second law of thermodynamics or Gisin's\cite{GISIN} demonstration of superluminal telephony. The
reason is that both of these arguments depend on assumptions made about linear behavior on
density matrices which need not hold in DRT's.
\vskip.1in
Following pioneering work by  Bohm and Bub\cite{BOHM}, the most successful DRT's to date are the spontaneous 
localization theories of  Ghirardi et al\cite{{GHIRARDIA,GHIRARDIB}}, and Pearle\cite{PEARLEB}. They are, however, ad hoc and
phenomenological, requiring the introduction of new constants of nature. Predictions are made concerning the spontaneous ejection
of electrons from atoms with resulting x-ray generation, and 
experimental constraints have been put
on the parameters\cite{COLL}. 
\vskip.1in
In this paper we shall develop a DRT differing fundamentally from spontaneous localization theories. The differences
will be found in three basic questions the theory must answer:
\vskip.1in
(a) The trigger: What initiates the collapse?
\vskip.1in
(b) The preferred basis: What determines the Hilbert space directions along which 
 collapse occurs?
\vskip.1in
(c) The noise: What produces it, and what is its fluctuation pattern? 
\vskip.1in
Our point of departure from  spontaneous localization theories is the observation 
that the only macroscopically distinguishable states whose collapse we have to explain are those that we know how to produce,
or at least can conceive of producing in a gedanken experiment. Since all such experiments involve  {\it entanglement},
we need only insure that the collapse mechanism apply to entangled states.
 Thus rather than postulating one universal collapsing interaction (such as the ``hits"
of spontaneous localization theory") that operates on all states, we shall adopt the following hypothesis:
\vskip.1in
 {\it Any Hamiltonian that can distinguish the factorized constituents of an entangled state induces a
corresponding non-linear interaction that can collapse the state into those constituents.}
\vskip.1in
 The strength and preferred basis of the
non-linear term will be inherited from the Hamiltonian that induces it. Since the theory will now have no
new constants of nature to set scales of length and time, it will be up to the dynamics itself to insure that
only macroscopically distinguishable superpositions collapse rapidly.
 We shall refer to this type of theory
as an {\it induced non-linearity}  (INL) theory. 
\vskip.1in
By allowing ordinary Hamiltonians to induce non-linear terms we are, of course, exposing the theory to 
 rejection if a conflict is discovered with the well-established and very accurate predictions of standard quantum
mechanics. The INL theory will be  partially protected by its modesty, i.e.\ the non-linear term has no effect on
factorized states.  The danger to the theory as well as its opportunities for new predictions
will  be found   where there is competition between the linear theory controlling the dynamics of an entangled state
and an induced non-linear term that may modify the dynamics. The following examples are intended to show that (1) such
situations are not encountered in ordinary applications of quantum mechanics, but that (2) interesting situations
nonetheless arise in  which the theory can be tested.
\vskip.1in
(1) The simplest type of measurement: 
\vskip.1in
Consider a Stern-Gerlach experiment in which one measures the spin of an electron in
an atom by means of an inhomogeneous magnetic field.  The entanglement of the spatial wave function of the atom
with the spin state of an electron is effected by the inhomogeneous field. If $\ket{\phi_o}$ denotes the
initial state, the transformation is:
$$\ket{\phi_o} \to \alpha\ket{\hbox{up}}\ket{\uparrow} +
\beta\ket{\hbox{down}}\ket{\downarrow}. \eqno{(1a)}$$
Here $\ket{\hbox{up,down}}$ indicate the time dependent spatial locations of the atom, and $\ket{\uparrow,\downarrow}$
indicate orthogonal spin states of the electron in the direction determined by the orientation of the magnet.
After sufficient time $\tau$, which depends inversely on the strength of the field, the spatial wave functions associated
with $\ket{\hbox{up,down}}$  no longer overlap and ultimately become sufficiently well separated that one 
can recognize the difference between them macroscopically. It is this separating action of the field on the
entangled state that recognizes the difference between its constituents, induces the non-linear term, and triggers the
collapse. The preferred
basis consists of the two states on the right of (1a). There is no significant competition between the collapsing mechanism
and the linear dynamics because the matrix exlements of the linear Hamiltonian between the spatially
separated constituents are already negligible when the non-linear term is active.
\vskip.1in
(2)  Mesoscopic states created with a Be$^+$ ion trap.\cite{WINELAND}: 
\vskip.1in
Although the interaction is
quite different, the situation is schematically similar to example (1). An initial zero point harmonic oscillator state
$\ket{\phi_o}$ of a single ion in the trap is entangled by a pair of Raman lasers. The $\ket{\hbox{up,down}}$ of
example (1) will now be classical-like coherent states, $\ket{\alpha e^{\pm i\phi/2}}$, involving gaussians
separated by as much as $80$nm. The 
$\ket{\uparrow,\downarrow}$ of example (1) will  be internal atomic states. Now, however,  $\phi$ is a  {\it controllable}
phase  through which one can sweep experimentally. Using a detection beam that resonantly couples $\ket{\downarrow}$ 
 to a an excited state producing ion fluorescence, one  measures  the probability $P_{\downarrow}(\phi)$ for the
internal state to be $\ket{\downarrow}$, and thereby induces collapse.
\vskip.1in
 As the experiment was recently described\cite{WINELAND}, the radiative linewidth of the fluorescence that
signals collapse is
$\approx 20 MHz$ and hence  the collapse process is so fast ($\approx 5 \times 10^{-8}s$) that, as in example (1), there is
no chance for the non-linear term to compete with the linear term, governing the internal dynamics of the entangled
state.  If it becomes possible to design the ion trap to speed up the evolution of the entangled state, or if one can
detect the internal state with a narrower fluorescent transition $\gamma$, the INL theory  (see below) predicts an observable
change in $P_{\downarrow}(\phi)$, namely a small displacement with respect to $\phi$ associated with a time-delay
$\hbar/\gamma$ in the collapse.
\vskip.1in
(3) A typical EPR experiment:
\vskip.1in
The right side of (1a) is replaced by a state
$\ket{\Phi}$ of two spin-1/2 particles, i.e.\ we have an entangled state  already formed
by some prior dynamical process.
When one of the partners, say particle-1, encounters a measuring device
the collapse process begins to operate immediately on the entangled state. The trigger
is the  Hamiltonian that recognizes the factorized constituents  by an
interaction with particle-1, and collapse must occur in the time $\tau$ 
required to distinguish the constituents. That Hamiltonian will determine 
a basis for particle-1 say $\ket{\uparrow,1},\ket{\downarrow,1}$. The state $\ket{\Phi}$
then has a unique representation in the form:
$$\ket{\Phi} = \alpha\ket{\uparrow,1}\ket{a,2} + \beta \ket{\downarrow,1}\ket{b,2} \eqno{(1b)}$$
in which the two states on the right are orthonormal. (The two states $\ket{a,2},\ket{b,2}$ are normalized but
not necessarily orthogonal.) The collapse now goes to either of the two states on the right with probability
$|\alpha|^2$ and $|\beta|^2$ respectively. Thus it is seen
that the preferred basis for the collapse is once again determined by the magnet and does not depend on
$\ket{\Phi}$. The collapse takes place when the two particles of the entangled state are far apart, so that
matrix elements of the linear Hamiltonian between the factorized states are negligible during the collapse.
Thus the linear Hamiltonian does not compete with the non-linear term.
\vskip.1in
The above examples show that under normal circumstances the collapse mechanism is either not there at all
because the state is factorized, or it dominates the linear dynamics to the extent that there is no
observable competition. Let us therefore try to look for situations where the linear dynamics induces
a {\it slow} process through which the factorized constituents are recognized, i.e.\ slow enough that the linear
dynamics has a chance to transform the entangled state significantly while the induced non-linear term is acting.
\vskip.1in
 This may happen, for example, {\it if the internal interaction has matrix elements
connecting the factorized constituents to distinct decay channels.} The following is a case in point that  will be
discussed in detail below:
\vskip.1in
(4) Decay of neutral K-mesons: 
\vskip.1in
In the statndard
model the CP eigenstates $\ket{K_{1,2}}$ are the entangled states
$$\ket{K_{1,2}} = 2^{-1/2}(\ket K_0 \pm \ket{\overline{K_0}}),\hbox{ in which }\;\ket{K_0} = \ket{s}\ket{\overline{d}},\quad
\ket{\overline{K_0}} = \ket{\overline{s}}\ket{d},\eqno{(2)}$$
where $s,d,\overline{s},\overline{d},s$ are quark and anti-quark flavors. In the absence of CP violation the states
$\ket{K_{1,2}}$ would be energy eigenstates of the weak interaction Hamiltonian, the strength of which can be measured by the
mass difference $\delta m$ between these two states. Now the weak interaction Hamiltonian is also able to
recognize the two factorized constituents $\ket{K_0},\ket{\overline{K_0}}$ through the semi-leptonic decay modes 
of $K_1$ or $K_2$. Here the sign of the charged lepton will determine whether the decay involved the conversion
of an $s$ or $\overline{s}$ quark. According to the INL theory this recognition of the factorized constituents
of an entangled state must induce a non-linear term that will drive a  collapse of the entangled state. The
strength of this effect will be $\hbar/\tau$ where $\tau$ is the time it takes for the distinction to be
recognizable. Since the branching ratio to semi-leptonic decay is negligible for $\ket{K_1}$ (where it is overwhelmed
by the two-pion decay mode), it is essentially the lifetime of $\ket{K_2}$ (for which the branching ratio is 66\% to the
semi-leptonic mode) that sets the time-scale. The INL theory then obtains an induced non-linear term whose
strength relative to the linear term is of order $0.001$. The competition between the linear
and induced non-linear term thus produces a small 
distortion of the $\ket{K_2}$ state so that the energy eigenstate, is not quite a CP eigentstate. In
this way the INL theory produces a small CP violation which, considering the coarseness of the model, is
surprisingly close to the experimental value.
\vskip.1in
One sees clearly in this example that the non-linearity is to be triggered by the
linear Hamiltonian which is also to fix the strength ($\hbar/\tau$) and preferred basis
($\ket{K_0},\ket{\overline{K_0}}$). Note that  the linear interaction now dominates the non-linear term. However, the
dynamics is such that the effect of the non-linear term shows up in a small but detectable way.
\vskip.1in 
As we remarked earlier, the main criticism that has been leveled at various proposed DRT's is that they
are ad hoc. To avoid this criticism we shall, in the remainder of this introductory section, set forth
the physical ideas that motivate the choices we will make below in constructing the INL theory.
\vskip.1in
The form of the noise:
\vskip.1in
 Whereas a linear Hamiltonian can
cycle an entangled state into a factorized state and back again, the collapse process is uni-directional.  This does not
mean that it is thermodynamically irreversible, however, for a pure state goes into a pure state, and there is no increase
in entropy. To produce uni-directionality  the dynamical transformation must be described
by an operator  whose domain is restricted to the entangled states, i.e.\ one that {\it becomes singular on factorized
states}.  The uni-directionality does not introduce an absolute arrow of time, but each measurement results in
the deflection of a ``pointer", which assigns a direction to time in an unpredictable way. This suggests that
there should be a random relative sign between the linear and non-linear terms, and so, if a convention is
adopted for the linear term, we should introduce a random  sign for the non-linear term.
 It is quite remarkable that such a simple form of noise will turn out
to reproduce the statistical distribution of outcomes predicted by the recipe of conventional quantum mechanics.
\vskip.1in
A much subtler problem has to do with the {\it intervals} in which the sign fluctuation is to occur. 
It is important that these fluctuations not occur so frequently as to slow down the collapse process. In
order to avoid introduction of a new scale-setting constant to define the interval, we must let the stochastic process itself
determine when these fluctuations are to occur.  In brief the idea is as follows: One can think of the noise fluctuation as
deciding whether one of two gamblers will win in a fair game of chance. In spontaneous localization theories  fluctuations
occur in such a way that the gamblers  are betting a random amount in every play until one of them loses his
entire fortune. This game, known as the ``gambler's ruin", can result in a game of long duration. However, when the noise is
merely a sign fluctuation, it is possible to play the game as ``double or nothing" with the stake in every play being that
of the player with the smaller remaining fortune. The average length of such a game is just two plays, and so, without having
to introduce a new time constant, the collapse  is not significantly delayed.
\vskip.1in
Form of the non-linear term:
\vskip.1in
Because the INL theory only collapses entangled states we can deduce another
 important property that the theory will be required to have.  Suppose we partition the Hilbert space into orthogonal
``cells", each of which is a connected set of states that can be collapsed by the non-linear map into the same set of
factorized states. If the non-linear map is to act independently on states belonging to different cells, it must be
extended by linearity to superpositions of states belonging to different cells. But this requires that the map be {\it
homogeneous}, i.e.\ it must retain a vestige of linearity, namely linearity with respect to multiplication by scalars.
\vskip.1in
We shall, of course, have to examine carefully what the map does on approaching the boundary of a cell.
In order to join the entangled state before collapse to a factorized state after collapse that satisfies the
ordinary \Schrodinger equation, the singularity on the cell boundary 
 will have to be weak enough that both a wave function and its time derivative change
continuously. Since quantum mechanical probabilities are computed from a quadratic form in the wave function, it follows
that probability must be conserved across the boundary. This suggests that the non-linear theory should
also retain another property of the linear theory, namely that  probability be conserved without the necessity
of having to renormalize wave functions.
\vskip.1in
As is suggested by example (3) above, there is an intimate connection between the INL theory and the
problem of EPR correlations. The INL theory is required to explain the rapidity of collapse of the wave
function asociated with an entangled state when a measurement is made. The existence of EPR correlations
reveals the most baffling aspect of the collapse process: If the measurement is made on one of the particles
of an entangled pair, the time required for the collapse is independent of the distance from the point of measurement to
the other particle and depends only on the reaction time of the measuring device. While this means, in particular,
that the collapse happens before there is time for a light signal to get from the point of measurement to the other
particle, the speed of light plays no intrinsic role in the EPR problem. The problem of quantum non-locality,
the violation of outcome independence\cite{SHIMONY}, is not going to be resolved by producing a Lorentz covariant 
form of the theory although that may be desirable for other reasons. 
\vskip.1in
There is, however, a type of symmetry that
will ``explain" what we observe and which we shall be able to implement in the INL theory: Suppose that
in the absence of two-particle interactions in the Hamiltonian, the INL theory can be constructed in such a way that its 
form is invariant under {\it arbitrary} unitary transformations of the form:
$$\Psi \to \Psi' = ( U^{(1)}\otimes U^{(2)})\Psi,\eqno{(3)}$$
i.e.\ any unitary transformation that acts separately on the two particles.  We shall say that such a theory  enjoys {\it
one-particle unitary invariance}. Since space-time symmetries in a quantum theory are implemented by 
such transformations, we see that one-particle unitary invariance guarantees that {\it the
INL theory will  preserve any space time symmetry enjoyed by the linear theory} whether it be Lorentzian or
Galilean. It also means that {\it the non-linear term will be unchanged in form if one applies
a translation operator to just one of the two particles, and hence the non-linear term will produce an effect that is
independent of the distance between the two particles.}
\vskip.1in
 By displaying a  collapse mechanism that is 
completely insensitive to the space-time symmetries of the linear theory, the INL theory essentially decouples it
from the notion of ``event" in space-time. Thus, in a sense, the INL theory dissolves the non-locality problem
by asserting that the notion of locality is meaningful only to the extent that the linear theory is 
a good approximation to dynamics.
\vskip.1in
We are now ready to embark upon the construction of the INL theory. The above discussion has supplied 
very strong constraints. We list them here for convenient reference. As we implement each of them
below the form of the theory will emerge:
\vskip.1in

({\bf A}) The non-linear interaction is uniquely determined  by a Hamiltonian that distinguishes 
the factorized constituents of
an entangled state.

({\bf B}) The non-linear map must be weakly singular on factorized states.

({\bf C}) The non-linear map must enjoy one-particle unitary invariance.

({\bf D}) The non-linear map must be homogeneous.

({\bf E}) The \Schrodinger equation with the non-linear term must conserve probability.

({\bf F}) The noise consists of a randomly fluctuating sign of the non-linear term.

({\bf G}) The occurrence of a fluctuation must be determined by the stochastic process itself.
\vskip.1in
Because these constraints 
suffice to essentially fix the form of the theory, and since we do not have to introduce any new constants, the INL
theory will neither be ad hoc nor phenomenological.
\vskip.2in 
\centerline{{\bf 2. The modified \Schrodinger equation.}}
\vskip.2in
Since entanglement plays a special role in the INL theory, we begin by introducing a simple formalism in which
the distinction between entangled and factorized states is clearly manifest. To keep the notation as simple as possible we
will develop the essential features of the theory for the simplest entangled system consisting of a pair of spin-1/2
particles in which only the spin degree of freedom is explicitly indicated. After developing the theory in this context the
generalization to more complex systems will be described.
\vskip.1in
The state of two spin-1/2 particles can be represented by:
$$|\Omega\rangle = C_{00}|0\rangle|0\rangle +
C_{01}|0\rangle|1\rangle +C_{10}|1\rangle|0\rangle +C_{11}|1\rangle|1\rangle \eqno{(4a)}$$
with the $C_{ij}$ assuming complex values. Dependence on space coordinates is not indicated.
 We represent the state $|\Omega\rangle$ by the
matrix of coefficients i.e.\
$$
|\Omega\rangle \to  C \equiv \left(\matrix{ C_{00} & C_{01}\cr C_{10} & C_{11}}\right), \eqno {(4b)}$$
so that scalar products become
$$\langle \Omega ' \ket{\Omega} = Tr(C'^{\dagger}C) ,\eqno{(5)}$$
and the normalization condition is:
$$\langle \Omega \ket{\Omega} = Tr(C^{\dagger}C) = 1.\eqno{(6)}$$
It follows from this (see Appendix A) that
$$ 0 \leq |\det C| \leq 1/2,\eqno{(7)}$$
and that $\det C = 0$ if and only if the state is a factorized state. At the other
extreme $|\det C| = 1/2$ corresponds to  completely entangled states.
\vskip.1in
Operators that act on the spin of particle-1 act on $C$ from the left, and those that act on particle-2
act on $C$ from the right. Thus the most general unitary transformations that act on the particles
separately are of the form
$$ C \to ACB, \eqno{(8)}$$
in which $A,B$ are two-by-two unitary matrices. Under Lorentz transformation, for example,
 $C$ is transformed
into $ACB$ in which  $A$ and $B$
 are the Wigner rotations associated with the
Lorentz transformation. These may be different for the two particles and are momentum dependent.
Since $|\det(ACB)| = |\det C|$ for
unitary
$A,B$ it follows that the properties of being a factorized
 or  maximally entangled state are unitary invariant properties.
\vskip.1in
In this language the ordinary \Schrodinger equation of the the system $S$ is:
$$
{dC\over{dt}} =  - iH_1 C - iCH_2 + R_{12}(C)
\eqno{(9)}$$
where $H_1$ and $H_2$ are the one-particle Hamiltonians, and $R$ is a linear, two particle interaction, e.g.\ a
spin-spin interaction (see eq. (35)). Thus the $H_j$'s contain the kinetic terms and any interactions which the particle
spins may experience with an external field. In the typical EPR experiment there will be no $R$ term and one of the $H_j$'s
contains the interaction with a Stern-Gerlach magnet which measures the spin of the particle. 
\vskip.1in
To construct a dynamical reduction theory  we add a term, writing
$$
{dC\over{dt}} = {\cal M(C)} - iH_1 C - iCH_2 + R_{12}(C).
\eqno{(10)}$$
We refer to the added term as the non-linear term and the remaining terms as the Hamiltonian or linear terms.
Our task is to find an appropriate form for the non-linear term.
\vskip.1in
Observe first that ${\cal M}(C)$ must be singular on factorized states in order to avoid recycling a factorized into
an entangled state. Now we note that $C \to C^{-1}$ has the right sort of singularity, since the factorized states
are characterized by $\det C = 0$. However when spins are transformed by unitary transformations we have $C \to ACB$ but
$C^{-1}
\to  B^{-1}C^{-1}A^{-1}$. We can, however, arrange to have the requisite singularity and still have the linear and
non-linear terms transform in the same way if we use the map $C \to C^{\dagger -1}$ where $\dagger$ is the hermitian
conjugate. In fact if $\nu$ is any real number we can take the map to be:
${\cal M}(C) =  |\det C|^\nu C^{\dagger - 1}, $ and we will then have ${\cal M}(ACB) = A{\cal M}(C)B \;$.
 But we saw above that the map must be homogeneous, and it is easy to check that for $n\times n$ matrices 
this requires $\nu = 2/n$. Thus for the present  situation with $n=2$ we are
 led to the  map:
$$ C \to \widehat{C} \equiv |\det C|C^{\dagger -1}.\eqno{(11)}$$
This map has remarkable properties.
\vskip.1in
(i) On its domain, which consists of non-singular matrices, $C \to \widehat{C}$  is homogeneous,
 maps unitary matrices into themselves, and is an {\it automorphism} of the
group of non-singular matrices, i.e.\ 
$$\widehat{C_1C_2} = \widehat{C_1}\widehat{C_2},\hbox{ if } \det C_1 \neq 0,\; \det C_2 \neq 0. \eqno{(12a)}.$$
so that in particular
$$ \widehat{ACB} = A\widehat{C} B,\; \hbox{ for unitary }\, A,B\eqno{(12b)}$$
This is the indispensible property needed to insure the one-particle unitary invariance of the theory to which we
referred in the introduction.
\vskip.1in
(ii) Consider any set ${\cal S}$ of entangled states
that have the same so-called ``Schmidt normal form", i.e.\ states of the form
$\gamma_1\ket{0}\ket{0'} + \gamma_2  \ket{1}\ket{1'}$ with$\gamma_1 \neq 0$ and $\gamma_2 \neq 0$  in which
 $\ket{0},\ket{1}$ and $\ket{0'},\ket{1'}$ are arbitrary orthonormal bases for the two particles.
Then  $C \to \widehat{C}$ leaves the set ${\cal S}$ invariant. To see this observe that the states of ${\cal S}$
are of the form $C = U\Gamma V$ where $U$ and $V$ are {\it fixed} unitary martrices (determined by the bases)
and $\Gamma$ ranges over all diagonal matrices such that neither of its diagonal elements $\gamma_1,\gamma_2$ vanish.
The asserted invariance property follows from (12b), i.e.\ $\widehat{U\Gamma V} = U\widehat{\Gamma} V \;$,
and a direct calculation showing that $\widehat{\Gamma}$ is diagonal and has no vanishing diagonal elements if this is true
for $\Gamma$.  The significance of this property will become clear below. It assures that the collapse takes place
within sets of the form ${\cal S}$ which we shall call ``cells". The {\it homogeneity} of the map will then permit the
extension of the non-linear map to linear combinations of states belonging to different cells. For example the
state on the right side of (1b) can be written as such a linear combination by expressing $\ket{a,2},\ket{b,2}$
in terms of any orthogonal basis and the predictions of the theory will be independent of the choice of that basis.
\vskip.1in
(iii) $C \to \widehat{C}$ is a {\it duality}, i.e.
$$ \widehat{\widehat{C}} = C.\eqno{(13)}$$
and is related in a simple way to time-reversal. To see this recall that the 
time-reversal operator ${\cal T}$ is an anti-unitary
map that can be defined on spin states by:
$$
{\cal T}(\lambda\ket{0} + \mu\ket{1}) = \lambda^*\ket{1} - \mu^*\ket{0}\eqno{(14)}$$
from which we deduce
$$
\widehat{C} = e^{-i\arg(\det C)}{\cal T}(C).\eqno{(15)}$$
Since ${\cal T}$ is non-singular we see that the singularity of the map arises from the phase factor and is merely an
{\it ambiguity} in the phase when the determinant goes to zero. This is a very important conclusion because, as we will see,
such a mild singularity causes no problems when we try to match up solutions across the domain boundaries of the map.
\vskip.1in
We must now put in the dependence on the Hamiltonian that induces the non-linear term. Its task is to define a basis along
which the collapse takes place and provide the energy distinguishing the different collapsed states. Let us try to do this
in the simplest possible way and guess the form of our modified \Schrodinger equation to be:
$$
dC/dt = \Lambda_1\widehat{C} + \widehat{C}\Lambda_2  - iH_1C - iCH_2 + R_{12}(C),
\eqno{(16)}$$
in which the matrices $\Lambda_j$'s are going to be determined by the inducing Hamiltonian.
\vskip.1in
Conservation of probability now imposes a strong constraint.  Thus from (6) we must have:
 $${d\over dt} \hbox{Tr}(C^\dagger C) = 0.\eqno{(17)}$$
This will be satisfied if the $\Lambda_j$'s  are hermitian and 
$$ \hbox{ Tr}(\Lambda_1 + \Lambda_2) = 0\eqno{(18)}.$$
\vskip.1in
Next consider the typical EPR experiment with a Stern-Gerlach magnet in which the linear dynamics recoginzes the
factorized constituents by acting on particle-1. Thus we set $\Lambda_2$ = 0. Since $\Lambda_1$ is hermitian
and has zero trace by (18) we can choose a basis for particle-1 in which $\Lambda_1$ is diagonal and of the form:
$$\Lambda_1 = \eta/2 \left(\matrix{1 & 0\cr 0 & -1}\right),\eqno{(19)}$$
It is clear that the parameter $\eta$ which has the dimensions of energy,
should be $\hbar/\tau$ where $\tau$ is a suitable measure of the time it takes the magnet to resolve the spins.
A dimensionless factor of order unity can, of course, be introduced that would have to be determined by
experiment.
\vskip.1in
To test this hypothesis and to obtain guidance on introducing the noise we turn in the next section
to an examination of the solution of (16) with this form of $\Lambda_1$.
\vskip.2in
\centerline{{\bf 4. Solution of the equation with a non-linear term only.}}
\vskip.1in
 Consider the case where  $R_{12}(C) = 0$.
It follows from the properties (12) that we may, as in the linear \Schrodinger equation, transform away the
one-particle linear terms in (16)   by going to the interaction picture, i.e.
$$C \to  e^{-iH_1 t}C e^{-iH_2 t}.\eqno{(20)}$$
\vskip.1in
Let us further assume that we only make a measurement on particle-1 which we represent by (19). 
Equation (16) now has the form 
$$
dC/dt = \Lambda_1 \widehat{C},\eqno{(21)}$$
subject to the requirement  $\det C \neq 0$ defining the domain of the map.
\vskip.1in 
 Now note that the right side  transforms states of the form
 $\gamma_1\ket{0}\ket{0} + \gamma_2\ket{1}\ket{1}$ into one another (diagonal $C$) and also transforms states of the form
$\gamma_1\ket{0}\ket{1} + \gamma_2\ket{1}\ket{0}$ into one another (anti-diagonal $C$). An arbitrary initial state
can be decomposed into a linear combination of these two types. From the homogeneity of the map we can then
solve (21) for the two cases separately and combine the results by linearity. 
  Since the equation has the same form for both we now examine  (21) when the initial $C$ is diagonal. 
Let
$$
y_0 = |C_{00}|^2,\; y_1 = |C_{11}|^2,\eqno{(22)}$$
which are the probabilities at any time of finding the system in the state $\ket{0}\ket{0}$ and $\ket{1}\ket{1}$, respectively.
We take the initial values to be 
$$ y_0(0) = \alpha, \; y_1(0) = 1 - \alpha.\eqno{(23)}$$
We then obtain from (21) with (11) and (19):
$$
{dy_0\over{dt}} = \eta \sqrt{y_o y_1},\qquad {dy_{1}\over{dt}} = - \eta\sqrt{y_oy_1}.
\eqno{(24)}$$
Let us extract the solution by a method that will readily generalize when we disccuss higher spin. Define
a dimensionless variable $\tau$:
$$
d\tau = |\eta|\sqrt{y_o y_1} dt.\eqno{(25)}.$$
The solution now has a different form for $\eta > 0$ and $\eta < 0$. For $\eta > 0$ we have
$$
y_0 = \alpha + \tau,\; y_1 = 1 - \alpha - \tau,\eqno{(26)}$$
whence for $0 \leq \tau < 1 - \alpha$ we have
$$
|\eta| t = \int_0^\tau d\tau [(\alpha + \tau)(1 - \alpha - \tau)]^{-1/2}
=  \arcsin(1 - 2 \alpha) - \arcsin(1 - 2 \alpha - 2\tau).\eqno{(27)}$$
from which:
$$ \tau = {1 \over 2}\{1 - 2 \alpha + \sin(|\eta| t - \arcsin(1 - 2 \alpha))\}.\eqno{(28)}$$
For $\tau = 1 - \alpha$, i.e.\ for
$$ t =  t_0(\alpha) = |\eta|^{-1}(\pi/2 + \arcsin(1 - 2 \alpha)),\eqno{(29)}$$
one sees from (26) that the right sides of (24) vanish, and {\it the process terminates with
$y_0$ having the value $1$ and $y_1$ having the value 0}. One may
check that $y_0$ and $dy_0/dt$ are continuous at $t = t_0$, but that $d^2y_0/dt^2$ changes
discontinuously from $(\eta/2)^2$ to zero at this point. This is a manifestation
of the singularity of the mapping upon arrival at the factorized states. But, as we have anticipated,
the singularity is so mild that we have the required continuity in the function and its first derivative
that enables us to extrapolate across the temporal boundary, joining the factorized state to
a solution of the linear \Schrodinger equation.
\vskip.1in
Next observe that if the sign of $\eta$ is reversed one obtains the same solution as in 
(26) but with the sign of $\tau$ reversed. Thus the process will now terminate when $\tau = \alpha$
and the factorized state will have $y_o = 0$ and $y_1 = 1$. If the initial state is completely entangled, i.e.\
if $\alpha = 1/2$, the factorization in both cases requires time 
$$
t_0 = {\pi\over{2\eta}}.\eqno{(30)}
$$
{\it Thus the time scale for completion of the process is determined by the parameter $\eta$ of the measuring
device and has nothing to do with the time required for a signal to get from one particle to the other.}
\vskip.2in
\centerline{{\bf  5. Form of the noise.}}
\vskip.2in
We are now ready to put in the noise and turn (16) into a stochastic differential equation. The solution obtained
above immediately suggests a simple and natural way to do this: We observe that a fluctuating sign of the parameter
$\eta$ will alter the outcome. Moreover this has an intuitive physical justification:  From (15) the map $ C \to \widehat{C}$
is a  scalar multiple of the time-reversal operator. Measuring devices break time reversal symmetry by having
a pointer deflect in one direction rather than another.  Thus a randomly fluctuating sign of $\eta$ expresses the
ambiguity in the direction of time until a measuremnt is registered. 
\vskip.1in
But there are many possible hypotheses we can make about the {\it intervals} in which the random sign fluctuations occur: are they
finite or infinitesimal, sporadic or regular?  To decide among these we can use two basic tests that  must be passed by a
satisfactory model: 
(i) The predictions of quantum mechanics must be recovered. (ii) The delay in producing the factorization because of
oscillations in the sign must not be such as to destroy the rapidity of achieving factorization, i.e.\ the time scale should
remain of order
$1/\eta$.
 \vskip.1in
It is convenient to consider  the noise as a function of 
the parameter $\tau$ in (25) (which is invariant under Lorentz boosts). The simplest choice  would be to let the sign
fluctuate randomly in any interval $\delta\tau$. Let us see that this will pass test (i) but will not pass test (ii):
\vskip.1in
In this model, according to (26),  $y_0$ increases and $y_1$ decreases by the same amount $\delta\tau$ or vice versa
according to the sign choice. If this fluctuates randomly 
 we will generate a random-walk with boundary problem (sometimes called the ``gambler's ruin"),
\cite{FELLER} i.e.\ one
in which the walk terminates when $y_0$ reaches either $0$ or $1$. In our situation the
 probability of a left move and a right move in the $\tau$ variable are equal. Suppose that $y_o$ is an integer multiple
of $\delta\tau$.  
 Then if $p(y_0)$ is the probability of ending with $y_0 = 1$ and $y_1 = 0$ starting from
some given value $y_0$ one sees that:
 $$p(y_0) = {1\over 2}p(y_0 + \delta\tau) + {1\over 2}p(y_0 - \delta\tau)\eqno{(31)}$$
 which implies that $p$ is linear in its
argument. Since we must have $p(0) = 0,\; p(1) = 1$  we obtain $p(y_0) = y_0$.
 \vskip.1in
Thus for randomly fluctuating sign we deduce that 
if $y_0$ starts off with value $\alpha$, the probability that
the process will terminate  with $y_0 = 1$ is $\alpha$. {\it But this is just the square of the amplitude of the
final state in the initial state and so  is just the prediction made by conventional quantum mechanics.}
\vskip.1in
But let us now see that  a difficulty with this model is revealed when we ask for the {\it duration} of the process.
The fluctuating sign means that the factorization is delayed. If we divide up the $\tau$ interval $[0,1]$
into segments of size $\delta\tau$, then,
starting from $y_o = \alpha$, one can show\cite{FELLER} that the expected number of moves required to end the process will
 be $\alpha(1 - \alpha)/(\delta\tau)^2$. But from (25) the time to traverse an interval $\delta\tau$ is linear in $d\tau$ and
hence an arbitrarily small subdivision will sooner or later delay the factorization to the point that
(30) is no longer a reliable estimate.
\vskip.1in
To overcome this difficulty we must have a fluctuation scheme that does not become arbitrarily rapid. However
there is no natural choice of sub-interval in $\tau$. Indeed, since we had to assume that $y_o$ was an integer multiple
of $\delta\tau$ we would have to let $\delta\tau$ become arbitrarily small.
{\it A way out is to allow a fluctuation at those times that
are singled out by the process itself.} Thus suppose we think of  $y_o$ and $y_1$ as the fortunes of two gamblers
who play ``double-or-nothing" with the bet in each play being the current fortune of the player who has less. Let a sign fluctuation
of $\eta$ take place whenever there is a play and let $y_o$ win if $\eta$ is positive and lose if it is negative. In this
model of the noise the fluctuations are finite but not sporadic. Let us see that both tests are now passed:
\vskip.1in
First observe that for $z \leq 1/2$ the probability $p(z)$ of obtaining $y_o = 1$ will satisfy
 $$
p(z) =  {1\over 2} p(2z)\eqno{(32a)}$$
which has the solution $p(z) = z $ as before. Thus test (i) is passed. But now the average number of moves required to
end the game is
$$1/2 + 2(1/4) + 3(1/8) + \cdots = 2, \eqno{(32b)}$$
so the expected time for termination is merely doubled from (30). Thus the mean collapse time is now given by
$$
t_0 = \pi/\eta \eqno{(33)},
$$
and we see that test (ii) is also passed.
\vskip.1in
We now have a theory of EPR correlations in systems of two spin-1/2 particles that reproduces the quantum mechanical
predictions and happens in a time characteristic of the measuring device
\vskip.4in
\vfill
\centerline{{\bf 6. Competition between linear and non-linear terms}}
\vskip.2in
A major goal of the theory is universality, i.e.\ that the non-linear terms be associated with linear terms that induce them.
This suggests the idea that we look for situations in which both are present and compete. The usual EPR experiment is not
of this type, for the Hamiltonian term associated with the magnet merely displaces an electron. Suppose, however, that
a non-linear term is induced by a spin-spin term in the Hamiltonian. This cycles the spin states against the action of
the non-linear term that is driving towards the factorized state. Let us see how the competition plays out: 
\vskip.1in
We write (16) in the form 
$$
{dC \over{dt}} = \Lambda\widehat{C}  + R_{12}(C)\eqno{(34)},$$ 
with
$$
 (R(C))_{jk} = \sum_{m,n = 0}^1 R_{jkmn}C_{mn}.\eqno{(35)}$$
To model the spin-spin interaction we only need a non-vanishing value denoted $\gamma$ for the two $R_{jkmn}$ that exchange $C_{00}
\leftrightarrow C_{11}$, and we take $\Lambda$ to be of the form (20). We then obtain a pair of coupled differential
equations for $C_{00}$ and $C_{11}$ from (16):
$$
{{dC_{00}}\over{dt}} = {1\over 2}\left(\eta { {|C_{00}C_{11}|}\over{ C_{00}^*}} + i\gamma C_{11}\right) $$
$$
{{dC_{11}}\over{dt}} = {1\over 2}\left(- \eta { {|C_{00}C_{11}|}\over{ C_{11}^*}} - i\gamma C_{00}\right) ,\eqno{(36)}$$
We shall solve (36) with $\eta > 0$ in the region where the non-linear term is defined, i.e.\ where $\det C
\neq 0$, so that neither
$C_{00}$ nor $C_{11}$ vanish.
\vskip.1in
It is straightformward to obtain equations for $d|C_{00}|/dt$ and for $d|C_{11}|/dt$ from (36) and thereby to
obtain an equation for $d|C_{00}C_{11}|/dt$. One can also obtain an equation for $d(C_{00}C_{11}^*)/dt$
and thence for the relative phase of $C_{00}$ and $C_{11}$. The result is the following. Let
$$
C_{00} = \cos(\theta/2),\; C_{11} = \sin(\theta/2)e^{i\phi},\;
\eqno{(37)}$$
so that $\{C_{00},C_{11}\}$  is represented by a point on a sphere with polar latitude $\theta$ and azimuth $\phi$.
We choose units in which $\gamma = 1$ so that $|\eta| >> 1$ is the regime of macroscopic measurement.  
After some algebra we obtain from (36):
$$d\theta/dt = -\eta +  \sin\phi ,\quad d\phi/dt =  \cos\phi \cot\,\theta.\eqno{(38)}$$
We then deduce that there is an integral of the motion involving $\theta$ and $\phi$:
$$\sin\theta\cos\phi(\tan(\pi/4 + \phi/2))^\eta \; = \hbox{constant}.\eqno{(39)}$$
\vskip.1in
Let us examine this solution in various regions: $|\eta| >> 1$ means domination by the non-linear term and
$|\eta| << 1$ means domination by the linear term. First consider all cases in which 
 $|\eta| > 1$. 
One then sees from (38) that  $d\theta/dt$ is always has the same sign so that $\theta$ changes 
monotonically.  Moreover the slope has absolute value exceeding $|\eta - 1|$, whence for large $\eta$ the
factorization at $\sin\theta \to 0$ is achieved in times of order $1/\eta$ as we found for a macroscopic
measuring device.  In all cases for $|\eta| > 1$ factorization will ultimately come about. If $|\eta$ is not
too much bigger than unity, the effect will be a slow collapse.
\vskip.1in
A particularly interesting case of this kind may be realizable, 
as we noted in the introduction, as a result of recently developed techniques for studying mesoscopic entangled
states, the so-called ``\Schrodinger cat states" of trapped Be$^{+}$ ions\cite{WINELAND}.
The delay $\hbar/\eta$ predicted by the INL theory is just the inverse of the width
$\gamma$ of the fluorescent transition by
means of which the factorized constituent containing $\ket{\downarrow}$, is recognized. If $\phi(t)$ is the time-evolving
coherent state phase, the INL theory in first approximation implies that
the probability $P_{\downarrow}(\phi)$ should shift to  $P_{\downarrow}(\phi')$ where $\phi'(t) = \phi(t + \delta t)$ with
$\delta t$ inversely proportional to the width of the fluorescent transition.
\vskip.1in
It is interesting to observe from  (38) that in all situations with $|\eta| > 1$  
the phase $\phi$ will tend to the 
same value $ \hbox{sign}(\eta) \pi/2$ as one approaches the factorization at $\sin\theta = 0.$ This limit
is achieved smoothly, so that we can join the unfactored and factored solutions together smoothly  in time
across the operator domain boundary. For macroscopic $|\eta|$ the factor 
$(\tan(\pi/4 + \phi/2))^\eta$ in (39) forces $\phi$ to remain very nearly constant until 
the state is almost factorized. It then changes rapidly from whatever its initial value was to
 $\hbox{sign}(\eta) \pi/2$
\vskip.1in
Now we consider the cases in which
 $|\eta| \leq 1$. Suppose first that $\eta = 0$. Then the stationary states will be the the states
$$s_{\pm} = \textstyle{
{1\over\sqrt{2}} }(|0\rangle|0\rangle \pm |1\rangle|1\rangle),\; \eqno{(40)}$$
 This corresponds to
 $\theta = \pi/2$ and  $\phi = 0$ or $\phi = \pi$. Now with non-zero $\eta$
 we see from (38) that there is  a stationary
solution with $\theta = \pi/2$ and $\sin\phi =  \eta$. {\it Because the state is stationary it can never
reach a factorization boundary and so, in accordance with the INL theory, no noise fluctuation ever occurs.}
\vskip.1in
 Thus we find that the stochastic equation actually has a stationary solution in which $s_{\pm}$ are
modified to:
$$s_{\pm}' =\textstyle{ {1\over\sqrt{2}} }(|0\rangle|0\rangle \pm e^{i\arcsin(\eta)}|1\rangle|1\rangle).
\eqno{(41)}$$
The effect of the phase in the second term is to admix $s_{-}$ and $s_{+}$.
\vskip.1in
We shall refer to the phase $\arcsin{\eta}$ as the  ``non-linearly induced phase". 
\vskip.2in
\centerline{{\bf 7. Application of the theory to the neutral K meson system.}}
\vskip.2in
Let us now apply the model of the last section to the neutral K system. Equations in
the following are to be understood as valid up to normalization constants that are 
irrelevant to the conclusions and are left out. Higher orders in small quantities are also left out.
 In the absence of
CP violation the eigenstates of the weak interaction Hamiltonian are 
$K_{1,2} = K_0 \pm \overline{K_0} $. Because of the CP violation the actual eigenstates are
are $K_S,K_L$ which are not quite the same. One finds that:
$$K_L = K_2 + \epsilon K_1 = K_0 - (1-2\epsilon)\overline{K_0} = K_0 - e^{i\delta}\overline{K_0},\eqno{(42)}$$
in which the experimental value is
 $$\epsilon = 1.6 (1 + i) \times 10^{-3}, \eqno{(43)}$$
so that
$$ \delta_{experimental} = 2i(1+i)1.6\times 10^{-3}.\eqno{(44)}$$
\vskip.1in
Now let us apply the theory above. The spin-spin interaction plays the role of the weak interaction mass matrix
that splits the $K_1,K_2$ masses\cite{MOHAPATRA}. In our model that splitting is by $\gamma$ so we take it to be the
experimental $K_1,K_2$ mass difference which has imaginary part because the masses are unstable. Thus we set:
$$
\gamma = (1 + i) 3.5 \times 10^{-6} eV.\eqno{(45)}$$
The parameter $\eta$ according to the INL theory is determined using (33) from the time required to distinguish the
factorized states. In this case $K_0,\overline{K_0}$ are distinguished by means of the sign of the charged lepton in a
semi-leptonic decay which reveals whether  an $s$ or an $\bar{s}$ was converted.
 The semi-leptonic branching ratio is 66\% of $K_L$ and
negligible in $K_S$ decays so that we compute the collapse time $t_0$ required to recognize a factorized state from
the $K_L$ lifetime divided by $0.66$. Thus, we can compute $\eta$ (up to sign) from (33) and obtain
$$ \eta = \pm 2.6 \times 10^{-8}.\eqno{(46)}$$ Thus the non-linearly induced phase is found to be:
$$
\delta_{theory} = \arcsin(\eta/\gamma) = \pm 2i(1 + i) 1.9 \times 10^{-3}.
\eqno{(47)}$$
Considering the coarseness of the model it is quite surprising that we have gotten the experimental
result to within 20\%. 
\vskip.2in
\centerline{{\bf 8. The EPR problem.}}
\vskip.1in
Leaving out two-particle interactions one sees that in virtue of (13b) the INL theory predicated on equation (16)
has the same form if for any unitary transformations $A,B$ one makes the substitutions:
$$\Lambda_1 \to A \Lambda_1 A^{\dagger},\quad H_1 \to  A H_1 A^{\dagger},$$
$$ 
\Lambda_2 \to B^\dagger \Lambda_2 B,\quad H_2 \to  B^\dagger H_2 B,\hbox{ and  }\, C \to ACB.\eqno{(48)}$$
This means that the non-linear term will share whatever space-time symmetries are enjoyed by the linear theory.
It also means that if one only measures  particle-1, the equation is unaltered by any unitary transformation
applied to particle-2, in particular to any translation of particle-2.  
{\it This is just the observed EPR effect in which the
correlation appears in a time that is independent of the separation between the particles and depends only
on the response time of the measuring device.}
\vskip.1in
In examining  the puzzling behavior of EPR correlations,
 Mermin\cite{MERMIN} 
suggested that we might do well to  ask: ``What is it about the way we think about the
world that makes us so puzzled ?"
Since we have now seen that the ``spooky" behavior of EPR correlations is obtained in the INL theory from
the insensitivity of the non-linear term to the particlular form of space-time symmetry assumed, and
since this in turn emerges from the peculiar structure of the non-linear term, we might rephrase
Mermin's question as follows:
Can we find a different way to think about the transformations of wave functions that will
eliminate our puzzlement?
\vskip.1in
The way we usually think about the transformation of a wave function $\psi$ is through a differential equation,
i.e.\ we write:
$$ \psi \to \psi + dt (d\psi/dt),\eqno{(49)}$$
in which $d\psi/dt = f(\psi,t)$ and the form of $f$ determines the dynamics. Thus even if $f$
is non-linear we are thinking of changes in (52) to the wave function as 
{\it additive} changes.
\vskip.1in
But an interesting thing about our  $C$ matrices representing two particle systems above 
 is that they have a {\it multiplicative} structure as well as an
additive structure. In general that multiplicative structure may be useless because, unlike the additive
structure, inverses don't always exist. But within the sets of states that are driven to factorization by
the stochastic process, inverses both exist and by their existence distinguish the elements of a set from
its boundary.
\vskip.1in 
Let us see then what dynamics looks like if we reformulate it via the multiplicative property. Define a
``cell" as a connected set of states
which can be collapsed into any one of a given set of orthogonal, factorized states.
 Let $C_o$ be a state inside a
cell and $C$ a state on the boundary. With the trace norm (6) defining a metric we can, for any $\epsilon > 0$,
make a chain of states $C_o,C_1,\cdots,C_{n-1},C$ such that all but the last are in the cell and such that
the distance between each state and its successor is $\epsilon$. Now consider the operator 
$$Z_{j} = C_{j+1} C_j^{-1}\eqno{(50)}$$ which is well defined for $j=0,\cdots,n-1$ and which maps each $C_j$ of the sequence
to its successor. By letting $\epsilon$ become as small as we like we can define a continuous set of transformations
taking us from $C_o$ to the boundary. Note that these are not unitary, in general, but they nonetheless map
each $C_j$ to a matrix of the same norm. Moreover each $Z_j$ {\it acts only on particle-1} since it acts from 
the left. Indeed if we examine how the $Z_j$'s transform under one-particle unitary transformations $A,B$ applied to the
two particles we see that:
$$Z_j = C_{j+1}C_j^{-1} \to (AC_{j+1}B)(A C_j B)^{-1} = AZ_jA^{\dagger},\eqno{(51)}$$
which makes it plain that its properties are independent of how one describes particle-2.
 Equivalently, by operating from the right with
operators of the form $C_j^{-1}C_{j+1}$, we could also accomplish the same thing by acting only on particle-2.
\vskip.1in
 Once we recognize that   $C_o$ can be dynamically mapped to $C$ 
by a sequence of transformations $Z$ that act only on one of the particles, we see that an EPR correlation is induced {\it
without ``doing or sending " anything to the other one.} 
\vskip.1in
To descrbe this process let us use (16) to obtain a more natural way of characterizing the dynamics of collapse than
(16) itself. We have:
$$  C_{j+1}C_j^{-1} = I + (C_{j+1} - C_{j})C_j^{-1} \to  I + dt(dC/dt)C^{-1} \eqno{(52)}$$
so that from (16) with just the non-linear term we obtain:
$$ Z  = 1 +  \Lambda\,V(C)dt,\quad
 V(C) \equiv |\det C|(CC^{\dagger})^{-1}\eqno{(53)}$$
which essentially replaces the differential equation with a prescription for computing evolution directly
as a sequence of operations on particle-1.
\vskip.1in
Since the additive properties of wave functions are the useful ones for Hamiltonian processes, and the multiplicative
properties are the useful ones for collapse processes, one suspects that a more natural description of quantum mechanics
 will be obtained if
we embed Hilbert space in a {\it normed ring}, the
topology supplied by the trace norm metric, $\sqrt{Tr(CC^\dagger)}$. We conjecture that within such a framework the
problem of quantum non-locality will appear a great deal less mysterious.
\vskip.2in
\centerline{{\bf 9. Generalizing the theory}}
\vskip.2in
\centerline{{\bf a. Hilbert spaces of higher dimension.}}
\vskip.1in
Up to this point we have confined our analysis to two spin-1/2 particles. In order to create a general theory of
 we must now turn to an investigation of Hilbert spaces of arbtrary dimension $n$, and consider what
happens  when $n$ becomes very large  This is necessary e.g.\ in order to investigate the EPR effect in the
context that it was originally proposed by Einstein, Podolsky, and Rosen\cite{EPR} which involves coordinate and momentum
rather than spin wave functions. The Hilbert spaces will now be infinite dimensional , but can always be approximated
by spaces of very large but finite dimension\cite{FIVELA}.
\vskip.1in
 Referring
to the homogeneity requirement in the discussion  leading to (12) we note that for Hilbert spaces
of dimension $n$ we must re-define the map with the power $\nu = 2/n$ of the determinant:
$$
\widehat{C} = |\det C|^{2/n}C^{\dagger - 1}. \eqno{(54)}$$
It is then straightforward (See Appendix B) to generalize the arguments used for $n = 1/2$.  
Since we are primarily interested in  what happens for $n$ large, let us take advantage of a simplification
that occurs if the dimension is a power of two.  We can then think of the
measurement as a sequence of
$\log_2(n)$ measurements analagous to the spin-1/2 measurement in each of which  the  state is collapsed into one that
lies in a space with half the dimension of its predecessor. In the first step we write (20) as an $n\times n$ matrix in which
the units appearing in (20) are replaced by projectors onto a pair of conjugate subspaces of dimension $n/2$. One will then
find that the time required for the reduction of a completely entangled state in dimension $n$ to a completely entangled state
in dimension
$n/2$ is again given by (33). Repeating the process we find that the total reduction time to a simple factorized state is:
$$t = {\pi \over {\eta}}\log_2(n). \eqno{(55)}$$
Suppose then that we wish to apply the method above to an EPR experiment involving the coordinate and momentum wave
functions instead of spin. The number of phase space cells of volume $\hbar^3$ 
for a macroscopic measurement would be of order $10^{100}$. Since one coherent state per phase
space cell suffices to span the Hilbert space, it suffices
to consider a Hilbert space of dimension $n = 10^{100}$ for which the logarithmic factor is only $300$.
Therefore $300$ repetitions of the process above will reduce the wave function so that even for energy differences as small
as $1 eV$ the reduction time would be smaller than one picosecond.
\vskip.1in
While we have not investigated the limit $n \to \infty$, the determinantal factor in (48) that might cause problems
is formally unity in that limit. We thus have grounds to be optimistic that the infinite limit will be sensible.
\vskip.1in
\centerline{{\bf b. Multi-particle systems.}}
\vskip.1in
To  generalize the INL theory  to n-particles, the $C_{ij}$
will be replaced by $C_{i_1,i_2,\cdots,i_n}$. If all but two of the indices are held fixed we obtain
a two-particle matrix associated with some pair of particles.  The concepts of inverse,
adjoint,  determinant, and factorization are all applicable to this matrix, as is the question of whether a
linear term in the Hamiltonian induces a non-linear term with respect to this pair of variables.
One expects that like all systems of more than two particles the dynamics will be difficult.
\vskip.3in
\centerline{{\bf c. Lorentz invariance.}}
\vskip.1in
Because, as we have noted, the non-linear term  has one-particle unitary invariance, the need to make the theory explicitly
Lorentz covariant is more a luxury than a necessity. We expect that it will be possible for the follwoing reasons: 
We can replace the \Schrodinger equation by
a Tomonaga-Schwinger equation, letting the energy parameters on the right side be replaced by  energy densities. The
non-linear term transforms properly when the spins are transformed by Wigner rotation. The domain of the non-linear operator
is defined covariantly since if in one frame a set of states can be mapped by a device into the same factorized state, that
will be true in every frame. Moreover our prescription for the noise  has been defined covariantly. Thus, for  given initial
data, we have a prescription for computing the wave function on any space like surface, and for every process there is a
process of the same probability in any frame.
\vskip.2in
\centerline{{\bf 11. Comments}}
\vskip.1in
The most pressing need is to find definitive experimental tests. These are necessarily exotic because we must
have competition between the linear dynamics that entangles the state and the and non-linear dynamics that recognizes the
factorized constituents. When 
we have spatially well-separated wave functions it is relatively
 easy to recognize factorized constituents but difficult to have the linear dynamics alter the
form of the entangled state during the collapse process. When we have closely spaced
spatial wave functions, the linear dynamics influences the entangled state, but it is difficult to provide
a mechanism  that recongizes the factorized constituents. The ion trap technology
provides a mesoscopic regime in which lasers can be used to have both terms  compete, and the K-meson decay exploits
a fortuitous aspect of the weak-interaction dynamics whereby  the semi-leptonic decay mode acts
to recognize the constituents. 
\vskip.1in
To conclude let us test the INL theory against the desiderata listed by
Shimony\cite{SHIMONY} for a non-linear modification of quantum mechanics:
\vskip.1in
(a) It is not restricted to situations of measurement.
\vskip.1in
(b)When applied to macroscopically distinguishable states it produces
rapid collapse. There is no gestation of  \Schrodinger cats for unacceptable time durations.
\vskip.1in
(c) It reproduces quantum mechanical predictions where it ought to.
\vskip.1in
(d) Collapse happens in a finite time. There are no stochastic ``tails".
\vskip.1in
(e) One cannot send superluminal messages.
\vskip.1in
(f) The dynamics accounts for the occurrence of definite outcomes of measurements performed with actual 
apparatus. 
\vskip.1in
Professor Shimony lists two other disiderata of which the INL theory fulfils the important one: 
It is not explicity covariant but it 
 has appeared in his lifetime.
\vskip.1in 

\vskip.2in

The author wishes to thank J.\ R.\ Dorfman, T.\ Jacobson, H.\ Jawahery, R.\ Mohapatra,  A.\ Shimony, J.\ Sucher, and
C.\H.\ Woo for helpful conversations, suggestions, and criticism.

\vskip.15in
\centerline{Appendix A}
\vskip.2in
Any completely entangled state of two spin-1/2 particles can be written:
$$ |\Omega\rangle = {1\over{\sqrt{2}}}\sum_{i = 0}^1 |i\rangle A |i \rangle,$$
in which $A$ is an anti-unitary operator\cite{{FIVELB}}.
Thus for a completely entangled state $|\det\, C| = |\det A|/2 =
1/2$.  Since  $|\det \, C| =
|\det(UCV)|$  for any pair of unitary or anti-unitary transformations
$U$,$V$ we can choose a basis in which $C_{01} = 0$. Then  $\det\,C$ vanishes if and only
if one of the two coefficients $C_{00}$ or $C_{11}$ vanishes. But if either vanishes along with $C_{01}$ the
state factorizes. Conversely if the sate factorizes the determinant vanishes by direct calculation. Now note that
$|\det\,C|$ is majorized by $|C_{00}||C_{11}| +  |C_{01}||C_{10}|$. Each of the two terms  is bounded by one half
the sum of the squares of the factors, the upper bound being achieved only when the two are equal. Since the sum
of the four squared moduli is unity it follows that $|\det\,C|$ is bounded  by $1/2$  and achieves the upper bound
only if $|C_{00}| = |C_{11}|,|C_{01}| = |C_{10}|,$ and $|C_{00}|^2 + |C_{01}|^2 = 1/2$. Choosing a one particle basis
so that either $C_{00}$ or $C_{10}$ vanishes one obtains one of the familiar entangled states. 
\vskip.15in
\centerline{Appendix B}
\vskip.2in
For Hilbert space dimension $n$ we saw that to obtain the homogeneity property we must define the map
$C \to \widehat{C}$ by
$$\widehat{C} = |\det C|^{2/n}C^{\dagger - 1}.\eqno{(B1)}$$
Let $\pi_1$ and $\pi_2$ be projection operators into complementary subspaces of dimension $m$ and $n - m$.
A measurement that determines in which of these two subspaces a given state lies generalizes the
notion of a Stern-Gerlach filter for spin-1/2. The matrix $\Lambda$ given by (20) will now be replaced by
$$
\Lambda = \eta/2\left(\matrix{{1\over m}\pi_1 & 0 \cr 0 & {- 1\over {n - m}}\pi_2}\right),\eqno{(B2)}
$$
which has zero trace as required to conserve probability. 
\vskip.1in 
For an initial state with $C$ diagonal one puts:
 $y_j = |C_{jj}|^2,\; j=0,\cdots n-1$, and obtain from (16):
$${{dy_j}\over{dt}} =  \eta(y_0\cdots y_{n-1})^{1/n} \cases{ m^{-1} & $j=0,\cdots,m-1$ \cr
-(n-m)^{-1}, & $j = m,\cdots, n-1$\cr} .\eqno{(B2)}$$
For $\eta > 0$ put
$$d\tau = \eta(y_o\cdots y_{n-1})^{1/n}dt,\eqno{(B3)}$$
whence with initial values
$$y_j(0) = \alpha_j,\quad j=0,\cdots n-1,\qquad \sum_j\alpha_j = 1, \eqno{(B4)}  $$ one obtains 
$$y_j =  \cases{ \alpha_j + n\tau/m & $j=0,\cdots,m-1$ \cr
\alpha_j -n\tau/(n-m), & $j = m,\cdots, n-1$\cr}, \eqno{(B5)}$$
so that
$$
\eta t = \int_0^\tau  d\tau \{ \prod_{j=0}^{m-1}(\alpha_j + {n\tau\over m})\prod_{j={m}}^{n-1}(\alpha_j -
{n\tau\over{n-m}})\}^{-1/n}.
\eqno{(B6)}$$
which must be inverted to obtain $\tau$ in terms of $t$. One does not have a closed form analagous to (28) in general.
The process terminates at
 $$\tau = (1 - m/n)\min_{j\geq m}(\alpha_j).\eqno{(B7)}$$
Let us now focus on the special case of a completely entangled initial state, so that we  choose
$$
\alpha_j = 1/n, \; \forall n,\eqno{(B8)}$$
for which we have
$$
\eta t = 2\int_0^\tau  d\tau ({1 \over n} + {n\tau\over m})^{-m/n}({1 \over n} - {n\tau \over{n-m}})^{-1 + m/n} $$
The termination time $t_o$ is obtained  setting $\tau = 1 - m/n$ which after some algebra gives an
expression for $t_o$ in terms of a hypergeometric function namely:
$$
\eta t_o = 2(1-{m\over n})F(1,1,1+{m\over n},1-{m \over n})\eqno{(B9)}$$
Thus in particular for $m = 1$ we have:
$$\eta t_o = 2 (1 - {1\over n})F(1,1,1 + {1\over n}, 1 - {1 \over n}) \to n \hbox{  for } n \to \infty.
\eqno{(B10)}$$
whereas for $m = n/2$ we have 
$$\eta t_o = {{\pi}\over{2}}.\eqno{(B11)}$$

\end{document}